\documentclass[aps,pre,showpacs,showkeys,amssymb,amsmath,superscriptaddress,reprint]{revtex4-1}

\usepackage{amsmath}
\usepackage{graphicx}
\usepackage{color}

\usepackage{verbatim}
\usepackage{amssymb}
\usepackage{bm}
\usepackage{url,hyperref}
\usepackage[usenames,dvipsnames]{xcolor}

\hypersetup{colorlinks=true, linkcolor=blue, urlcolor=blue, citecolor=blue}
\usepackage{hhline}

\newcommand{\RM}[1]{\MakeUppercase{\romannumeral #1{}}}

\begin{document}

\title{Unveiling the Asymmetry in Density within the Shear Bands of Metallic Glasses}
\author{Harald R\"osner}
\email{rosner@uni-muenster.de}
\affiliation{University of M\"unster, Institute of Materials Physics, Wilhelm-Klemm-Str. 10, 48149 M\"unster, Germany}

\author{Arabinda Bera}
\affiliation{University of Milan, Department of Physics ``A. Pontremoli'', via Celoria 16, 20133 Milan, Italy}

\author{Alessio Zaccone}
\email{alessio.zaccone@unimi.it}
\affiliation{University of Milan, Department of Physics ``A. Pontremoli'', via Celoria 16, 20133 Milan, Italy}
 
\date{\today}
\begin{abstract}
Plastic deformation in metallic glasses at room temperature leads to the development of shear bands due to shear localization. In many experiments, shear bands have shown local density variations along their path, with a distinct imbalance in magnitude between local densification and dilation. However, a comprehensive mechanistic understanding or theory to explain this asymmetry has been lacking until now. Here, we introduce a new model that consists of a sequential arrangement of alternating topological 'charges',  generating a dipolar field. The resulting microscopic displacement field, when integrated into the deformation gradient tensor, provides an accurate analytical solution for the observed imbalances in the density variations. The implications of this method are discussed, highlighting the potential to elucidate a broader range of observations in shear bands.

\end{abstract}

\maketitle
\section{Introduction}
Metallic glasses (MGs) have been a prominent focus of materials research and condensed matter physics since their successful fabrication in 1960 \cite{klement1960non}, primarily due to their disordered structure \cite{wang2004bulk,schuh2007mechanical}. MGs exhibit remarkable properties such as high strength, fracture toughness, and a large elastic limit ($\epsilon_{el}$ $\approx$ 2 $\%$, much higher than the $\leq$ 0.1 $\%$ of their crystalline counterparts). They also possess unique soft magnetic properties, as well as distinct chemical and topological characteristics, suggesting wide-ranging applications \cite{wang2011plastic,bera2024soft}.

However, despite their high strength, metallic glasses have limited ductility and are prone to immediate catastrophic failure in tension after surpassing the elastic limit \cite{zhang2006making}. The plastic deformation of metallic glasses can be categorized into two regimes \cite{spaepen1977microscopic}: low temperature and high stress, and high temperature and low stress. In the former regime, deformation occurs within narrow zones known as shear bands, which are regions of intense shear strain and thus render the material mechanically unstable \cite{pampillo1975flow,pollard2022yielding}. 
Typically, shear bands have widths ranging from 5-20 nm \cite{zhang2006thickness,Rosner2014,schmidt2015quantitative,hieronymus2017shear,hilke2019influence}. The shear band core experiences a structural transformation caused by shear localization, which leads to shear dilation with a localized increase in volume or change in density \cite{donovan1981structure,falk1998dynamics,klaumunzer2011probing,greer2013shear,maass2015shear}. This alteration in volume or density leads to progressive softening of the shear band core, rendering it softer than the surrounding matrix \cite{bei2006softening,pan2011softening}. As a result, the applied shear strains can be accommodated through slip. Interestingly, 
while the overall free volume increases, local regions of densification may also occur \cite{donovan1981structure,Rosner2014}. However, there are also reports in literature where shear bands exhibit only dilation \cite{schuh2007mechanical,li2002nanometre,jiang2005mechanical,pauly2009crack,wilde2011nanocrystallization,liu2017shear,maass2020beyond,liu2021strain,rosner2022situ,kang2024importance}.
Therefore, measuring the excess free volume or density within shear bands with a high lateral resolution is of great significance. 
Various techniques, including atom probe tomography (APT) \cite{balachandran2019elemental,chellali2020deformation}, transmission electron microscopy (TEM) \cite{donovan1981structure,Rosner2014,schmidt2015quantitative,liu2017shear,hilke2019influence,a2020correlations,mu2021unveiling,sheng2022mapping,abrosimova2023changes,kang2023direct}, x-ray diffraction tomography \cite{scudino2023strain}, and simulation \cite{hassani2019probing,bamer2023molecular}, have been employed to investigate the local density or strain along the shear band path and  have shown variations between local densification and local dilation with a significant imbalance in magnitudes \cite{Rosner2014,schmidt2015quantitative,hilke2019influence,hassani2019probing,a2020correlations,sheng2022mapping,kang2023direct,scudino2023strain}.  

While our previous approach in Ref. \cite{hieronymus2017shear} provided a qualitative explanation for the density variation, it fell short in providing a quantitative solution for the observed asymmetry between local densification and dilation. In this paper, we present a deeper analysis of this physical phenomenon. In Sec. \RM{2}\textbf{A} we begin by scrutinizing the experimental observations, which are subsequently elucidated using continuum mechanics in Sec. \RM{2}\textbf{B}. Here, the key advancement arises from the use of the deformation gradient tensor, which leads to a novel analytical solution. This solution not only accurately captures the observed asymmetry in density variations, but also clarifies observations in shear bands which show no apparent densification, that is exclusive dilation. In Sec. \RM{2}\textbf{C} topological defects are used to recreate the periodic displacement field used in Sec. \RM{2}\textbf{B}. Finally, we discuss the implications of our model in Sec.  \RM{3} and conclude in Sec. \RM{4}.

\section{Results}
\subsection{Experimental Observations}
\begin{figure}[h]
\includegraphics[width=8.6 cm]
{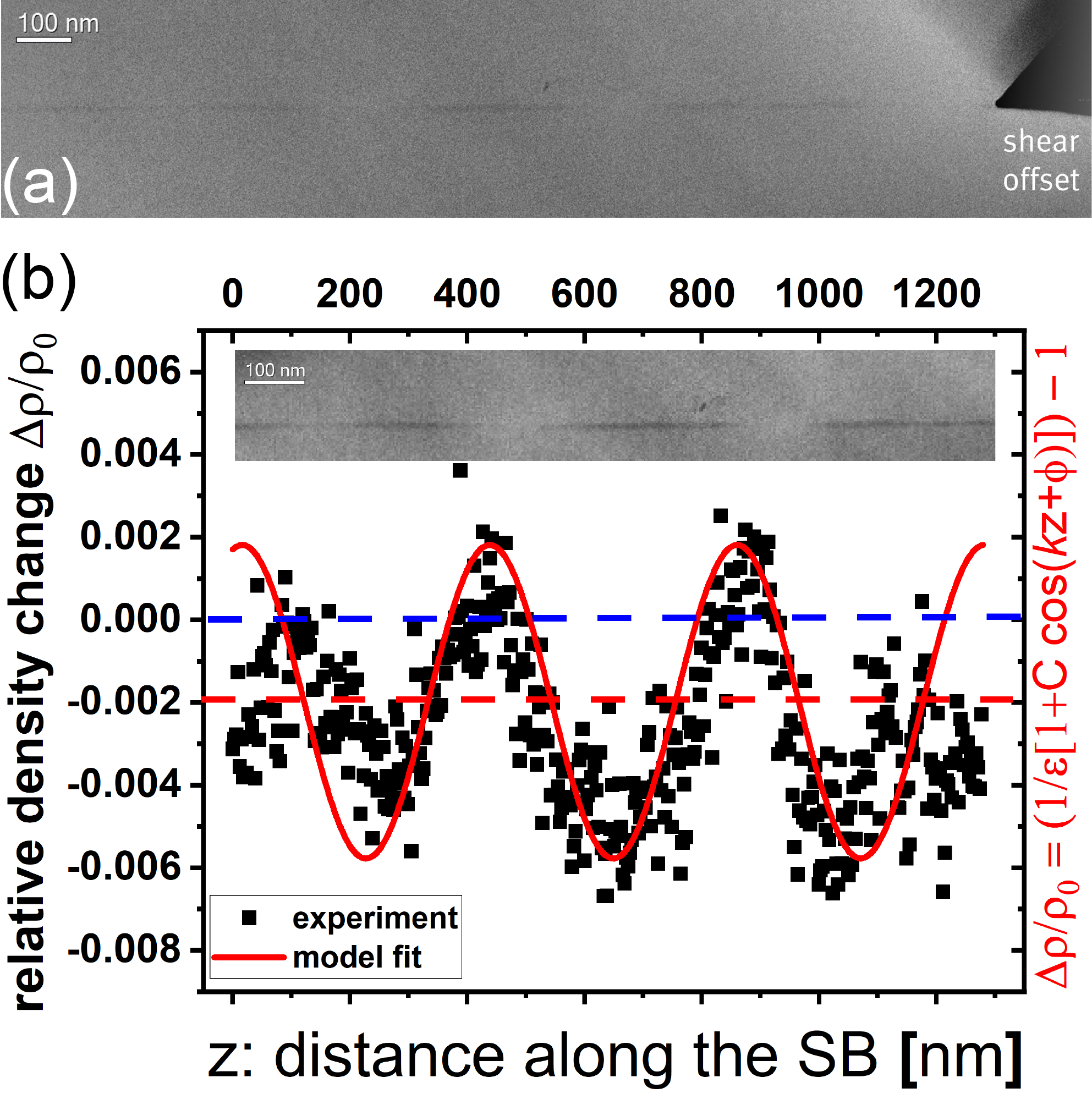}
\caption{(a) Z contrast image of a shear band in Vitreloy 105 (Zr$_{52.5}$Cu$_{17.9}$Ni$_{14.6}$Al$_{10}$Ti$_{5}$) exhibiting characteristic contrast variations along the path, ending in a shear step at the surface. (b) Inset: Contrast enhanced detail of the shear band shown in (a). Underneath: relative density changes extracted from the image (black data squares) and the analytical solution (red curve) fitted using Eq.~\eqref{model}. The blue line refers to the matrix as reference and the red line to the average relative density change in the shear band. Note the asymmetry between positive and negative magnitudes relative to the matrix.}

\label{Figure1}
\end{figure}

Fig.~\ref{Figure1}(a) displays part of a FIB lamella containing a representative shear band in a bulk metallic glass (Vitreloy 105) after deformation. This image was obtained through Z contrast imaging \cite{pennycook2002structure}. Due to the edge-on imaging conditions, the shear band appears as a narrow band which extends horizontally and ends in a shear offset at the surface of the foil. According to the reference coordinate system shown in Fig.~\ref{Figure2}(a), the $yz$ plane is the image plane, with $y$ perpendicular to the shear band. The $x$ direction is perpendicular to the image plane and the $xz$ plane corresponds to the shear band plane. Characteristic alternating contrast changes can be discerned along the shear band, i.e. along the $z$ direction. These contrast changes arise from differences in the density along the shear band and are not unique to Vitreloy 105 but have also been observed in various metallic glasses with different characteristic properties \cite{Rosner2014,schmidt2015quantitative,hieronymus2017shear,a2020correlations}. To analyze the observed variations in contrast along the shear band path, we use the intensity ratios of the high-angle annular dark-field scanning transmission electron microscopy (HAADF-STEM) signal. The conversion process works in the following manner: First, we extract a HAADF-STEM intensity profile from within the shear band along the propagation direction, as well as two profiles equidistant on each side. Averaging these latter two profiles determines the expected matrix intensity profile at the shear band position. Next, we subtract the matrix intensity profile from the shear band intensity profile and normalize the resulting difference by the matrix intensity profile. For a more detailed description of the density determination method, please refer to the references provided \cite{Rosner2014,hieronymus2017shear,hilke2019influence}. This procedure enables us to extract density changes within the shear band relative to the matrix. The resulting data are plotted as black squares in Fig.~\ref{Figure1}(b) and approximate to a sinusoidal curve. The analytical solution (red curve) was fitted based on the model solution described below in  Eq.~\eqref{model}. Note that the maximum positive density magnitudes are three times smaller than the maximum negative ones, showing an asymmetry in the density changes within the shear band. Thus, a macroscopic measurement of the shear band would result in a negative value for the relative density change as indicated by the red dashed line in Fig.~\ref{Figure1}(b), that is dilation \cite{klaumunzer2011probing,jiang2017shear}.

\subsection{Continuum Mechanics}
In Ref. \cite{hieronymus2017shear}, the mathematical analogy between elasticity theory and electromagnetism \cite{muggli1971analogy} was exploited to derive the following form of the displacement field $u$ along the shear band, resulting from an alignment of force dipoles:
\begin{equation}\label{displ}
    u(z) =\frac{A}{K}\sin (k z + \phi)
\end{equation}
where $K$ is the bulk modulus of the material, $z$ is the distance along the shear band, $k$ is the wavenumber, $\phi$ is an arbitrary phase, and $A$ is a prefactor. 

\begin{figure} 
\includegraphics[width=8.6 cm]
{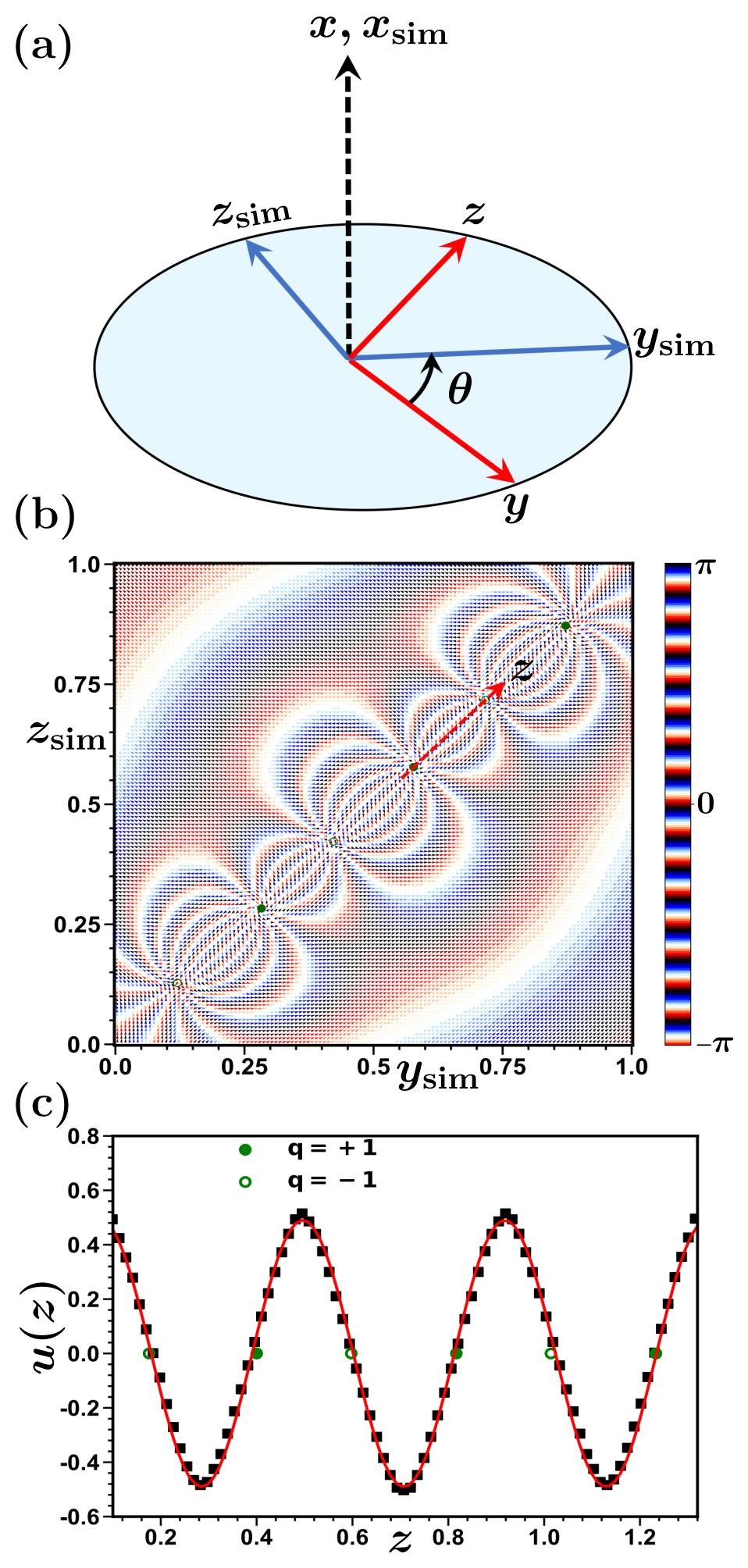}
\caption{(a) Schematic illustrating the reference coordinates associated with the experiment, called shear band frame $(x,y,z)$, and the simulation frame $(x_{\rm{sim}},y_{\rm{sim}},z_{\rm{sim}})$. The angle $\theta$ denotes the rotation angle between these two frames. (b) Sequential arrangement of topological charges with winding numbers $q=+1$ (filled circles) and $q=-1$ (open circles) within the shear plane ($y_{\rm{sim}}z_{\rm{sim}}$) of the simulation frame. The vector field reveals a dipolar field along the diagonal ($z$ direction), with the ``flag'' color bar indicating the phase angle. (c) Displacement field $u(z)$ extracted from (b) along the diagonal ($z$ direction), with the locations of charges having $q=+1$ and $q=-1$. The red line represents the fit of the numerical data points using $A'\sin(kz+\phi)$, with $A'=0.49$, $k=14.89$, and $\phi=0.46$.}

\label{Figure2}
\end{figure}

We utilize the 3D Cartesian reference frame [shown as $x, y, z$ in Fig.~\ref{Figure2}(a)] in which the shear band in Fig.~\ref{Figure1}(a) is aligned in the $z$ direction. 
The deformation process leading to the formation of shear bands is now mimicked by integrating the displacement field within the framework of the deformation gradient tensor. In continuum mechanics, the components of the deformation gradient tensor are defined as $F_{ij}=\delta_{ij} + \frac{\partial u_{i}}{\partial x_j}$, where $i,j$ denote Cartesian components, and $\delta_{ij}$ is the Kronecker delta. The corresponding Jacobian is defined as $J=det(\mathbf{F})=det(\delta_{ij} + \frac{\partial u_{i}}{\partial x_j})$. The volume of an infinitesimal element in the deformed sample is given by: $dV = det(\mathbf{F}) dV_{0}$, where $dV_{0}$ is the volume element in the undeformed sample \cite{tadmor}. Hence, $dV/dV_{0}=J=det(\mathbf{F})$. 
In the absence of chemical (mass) changes \cite{hilke2019influence}, the volume change may be expressed by the mass density change, $\Delta \rho$. Taking $\rho_{0}$ as the mass density in the matrix and $\rho$ as the mass density in the shear band gives \cite{tadmor}: $\rho/\rho_{0}=1/J$. We therefore have:
\begin{equation}\label{delta}
   \frac{\Delta \rho}{\rho_{0}}=\frac{1}{J}-1.
\end{equation}
The deformation gradient tensor may be written as:
\begin{equation}\label{tensor}
		\mathbf{F} =
 		\begin{pmatrix}
 		 1+\alpha & 0 & 0 \\
  		 0 & 1+\beta & 0 \\
  		0 & 0 & 1+\frac{\partial u(z)}{\partial z} \\
  		 \end{pmatrix} 
\end{equation}
where $\alpha$ and $\beta$ are both strain parameters. It is worth noting that although the deformation gradient tensor is presented in diagonal form, shear is present due to the non-zero difference between the diagonal components $F_{zz}$ and either $F_{xx}$ or $F_{yy}$.\\

Upon considering that, based on Eq.\eqref{displ},\\ $\frac{\partial u(z)}{\partial z} = C \cos(kz +\phi)$ with $C=k A/K$, we thus obtain
\begin{equation}\label{Jacobian}
    J = det(\mathbf{F})=\epsilon (1+C \cos(kz +\phi)) 
\end{equation}
with $\epsilon = (1+\alpha)(1+\beta)$ being the dilatancy in the shear band.
Upon substituting Eq.\eqref{Jacobian} into Eq.\eqref{delta}, for the density variation in the shear band relative to the matrix, we arrive at:
\begin{equation}
    \frac{\Delta \rho}{\rho_{0}}=\frac{1}{\epsilon (1+C \cos(kz +\phi))}-1.
\label{model}
\end{equation}

This equation was fitted to the  experimental data using the least squares approximation method and the resulting parameters are presented in Table~\ref{tab:TAB1}. The result of this fit is plotted in Fig.~\ref{Figure1}(b) as a red curve.

\begin{table}[h]
    \caption{List of fitting parameters.}
    \small 
    \centering
    \begin{tabular}{|p{4cm}|c|}
        \hline
        \multicolumn{1}{|c|}{\textbf{Fitting Parameter}} & \textbf{Zr$_{52.5}$Cu$_{17.9}$Ni$_{14.6}$Al$_{10}$Ti$_{5}$} \\
        \hline
        \multicolumn{1}{|c|}{$C=k A/K$} & $0.0038$ \\
        \hline
        {Dilatancy, $\epsilon = (1+\alpha)(1+\beta)$} \newline
         with $\alpha$, $\beta$ $>0$ 
         & $1.002$ \\
        \hline
        \multicolumn{1}{|c|}{Bulk modulus $K$ / [GPa]} & $113$ \cite{hieronymus2017shear} \\
        \hline
        \multicolumn{1}{|c|}{Wavenumber $k = \frac{2\pi}{\lambda}$ / [1/nm]} & $0.01489$ \\
        \hline
        \multicolumn{1}{|c|}{Wave length $\lambda$ / [nm]} & $422$ \\
        \hline
        \multicolumn{1}{|c|}{arbitrary phase $\phi$} & $2.9$ \\
        \hline
        \end{tabular}
    \label{tab:TAB1}
\end{table}

\subsection{Concept of Topological Charges}
Previous studies have linked mechanical failure and plastic responses in amorphous materials to quadrupolar (Eshelby-like) irreversible processes \cite{lemaitre2021anomalous,bhowmik2022direct,pollard2022yielding}. For a high quadrupole density \cite{shi2023dipolar}, the resulting field can be approximated as dipoles, with dipole-dipole and dipole-displacement forces being the primary interactions \cite{bhowmik2022direct}.
In general, these dipole arrangements introduce screening effects into the material behavior \cite{lemaitre2021anomalous,bhowmik2022direct}. Drawing inspiration from the electrostatics screening concept, akin to the Debye-H\"uckel approach \cite{lemaitre2021anomalous,landau} involving mobile charges (or monopoles), a similar framework is here extended to amorphous solids subjected to external shear. We inspect an arrangement of sequential topological defects \cite{wu2023topology,bera2024soft} carrying opposite 'charges' situated within the shear plane of the material. This sequential arrangement gives rise to dipolar fields and is used to generate a microscopic displacement field \cite{rouzaire2021defect}. Unlike the previous studies where dipoles were identified as gradients of quadrupole fields \cite{lemaitre2021anomalous,bhowmik2022direct}, we describe them as the resultant field emanating from two topological defects with opposite charges (vortex and anti-vortex) \cite{wu2023topology} (see also \cite{Sopu,Falk2024}). The structural pattern of the displacement field resulting from this construction exhibits a remarkable resemblance to the non-affine displacement field \cite{Scossa} of Eshelby inclusions aligned along a shear band \cite{gendelman,dasgupta2012microscopic}.

Topological defects manifest as singular points or lines within scalar, vector, or tensor fields, distinguished by fundamental topological invariants like the winding number or topological charge. The topological charge $q$ associated with a phase $\vartheta$ of a vector field in a two-dimensional plane is defined as $q = \frac{1}{2\pi} \oint_{\mathcal{L}} d\vartheta$ \cite{wu2023topology,rouzaire2021defect,missaoui2020annihilation,tang2017orientation}, where $\mathcal{L}$ is a closed loop enclosing a defect core. A topological charge of $q=+1$ denotes a vortex, while $q=-1$ signifies an anti-vortex. At the shear plane ($y_{\rm{sim}}z_{\rm{sim}}$) shown in Fig.~\ref{Figure2}(b), the phase angle $\vartheta_p$ linked to a vortex is expressed as $\vartheta_p = \pi/2 + {\tan}^{-1}[(z_i - z_p)/(y_i - y_p)]$, where $(y_p, z_p)$ indicates the vortex's location in the simulation frame. Conversely, the phase around an anti-vortex is characterized by $\vartheta_n = 2\pi - {\tan}^{-1}[(z_i - z_n)/(y_i - y_n)]$, with $(y_n, z_n)$ representing the anti-vortex's position \cite{rouzaire2021defect,missaoui2020annihilation,tang2017orientation}. The interactions and annihilation processes of these topological point defects with opposite charges in 2D and quasi-2D systems have been thoroughly investigated in several prior studies \cite{rouzaire2021defect,missaoui2020annihilation,tang2017orientation}.

In the presence of multiple topological defects within the system, the effective field arises from the superposition of phases originating from individual defects \cite{rouzaire2021defect,missaoui2020annihilation,tang2017orientation}. For the dipolar field, we compute the phase $\vartheta = \sum_{k=1}^N (\vartheta_{p,k} + \vartheta_{n,k})$, where $N$ denotes both the number of vortices and anti-vortices. The dipolar field [$\vec{u} = (\cos\vartheta, \sin\vartheta)$] is constructed with the locations of defects systematically arranged along the diagonal, as illustrated in Fig.~\ref{Figure2}(b).

The resulting field manifests variations along the dipolar field as depicted in Fig.~\ref{Figure2}(c) and correlates with the ansatz made in Eq.\eqref{displ}. The resultant field in the vicinity of the charges along the diagonal is characterized by $u(z) = A'\sin(kz + \phi)$, where $k = 2\pi/b$, and $b$ $(\approx 0.42)$ represents the distance between two identical charge defects. $A'$ and $\phi$ are constants with $A' \approx 0.49$ and $\phi \approx 0.46$. This distinctive concept of topological charge draws parallels with a Coulomb gas composed of particles featuring equal-magnitude positive and negative charges where the interactions are governed by the Coulomb interaction which depends on the distance between them \cite{minnhagen1987two}.

\section{Discussion}
The mathematical framework described in Eq.\eqref{model} provides an accurate solution aligning closely with the experimental findings. We now provide a physical interpretation by examining the reciprocal term ${\epsilon [1+(kA/K) \cos(kz +\phi)]}$ in Eq.\eqref{model}. The key parameters are $\epsilon$, $k$ and $K$. $\epsilon$ is the dilatancy caused by the shear and relates to the average relative density change, shown by the red dashed line in Fig.~\ref{Figure1}(b). A higher $\epsilon$ would shift this line further down. The periodic nature of the resultant function arises from the cosine term. The wavenumber $k$ is related to the periodicity of the function and hence to the separation of the defects. It also affects the amplitude of the sinusoidal function. The amplitude is also affected by the bulk modulus, $K$. A larger $K$ would result in a smaller amplitude of the sinusoidal function, as would a smaller wavenumber $k$. Evidence for this phenomenon has been observed in Al$_{88}$Y$_{7}$Fe$_{5}$ \cite{Rosner2014,schmidt2015quantitative}, a metallic glass exhibiting a significantly lower bulk modulus yet  demonstrating amplitudes over ten times larger when compared to Pd- or Zr-based bulk metallic glasses \cite{hieronymus2017shear,hilke2019influence, davani2020shear}. 
Thus, the interplay of both $k$ and $K$ dictates the  amplitude of the sinusoidal curve and $\epsilon$ its shift relative to the origin. A sufficiently large $\epsilon$ value could shift the red curve in Fig.~\ref{Figure1}(b) down such that the oscillations are solely in the negative range (dilation only). Such a case would be consistent with recent findings which show a trend of heightened local volume dilation as shear strain increases \cite{liu2017shear,liu2021strain}. It should also be noted that local strain relief occurs in thin electron-transparent TEM foils if a critical sample thickness is not maintained \cite{kang2024importance}. This could explain a substantial portion of observations, where no clear density oscillations were observed.

While our model provides a value for the dilatancy, $\epsilon$, it does not reveal the individual values for the strain parameters $\alpha$ and $\beta$. These values could be either equal or different. In the scenario where $\alpha$ and $\beta$ are equal, we can determine them from our data fitting in Table~\ref{tab:TAB1} to be approximately $0.001$, a value that aligns well with reported volumetric strain measurements within a shear band \cite{kang2023direct,scudino2023strain,kang2024importance}.

A further point of interest pertains to the topological defects involved in the shear band formation. The dipolar field depicted in Fig.~\ref{Figure2}(b) arises from the sequential arrangement of alternating defect charges, which can be classified as positive and negative topological defects \cite{wu2023topology}. The negative charge defects, resembling anti-vortices \cite{wu2023topology}, have been associated with the locations of plastic events triggering plastic flow, as documented in recent studies \cite{bera2024soft,Falk2024}. These defects mimic quadrupolar stress fields similar to those surrounding Eshelby's inclusions in an elastic continuum \cite{eshelby1959elastic}, which are believed to govern plasticity in amorphous materials \cite{lemaitre2021anomalous}. Evidence has recently been reported for the existence of Eshelby inclusions within shear bands \cite{kang2023direct}. The observed separation distances for the cores of these Eshelby inclusions fall within the range of ($375$ - $550$) nm, aligning well with the separation of the defects in our study, which measures 422 nm. While our new model presents a simpler and more generic microscopic alternative, it is, however, consistent with the previous paradigm based on Eshelby quadrupoles, where the dipolar fields stem from the gradient of the quadrupolar fields \cite{lemaitre2021anomalous,bhowmik2022direct}. The advantage of our new model lies primarily in the fact that no fictitious "inclusions'' have to be artificially postulated, which was necessary for the Eshelby quadrupoles. Furthermore, in the present model, the dipolar field arises naturally from topological defects of opposite charges (vortex and anti-vortex) \cite{wu2023topology}, which still allows us to take advantage of the analogy with electrostatics, as previously suggested in our work on shear banding \cite{hieronymus2017shear}. As topological defects originate from non-affine displacements of atoms \cite{baggioli2021plasticity,baggioli2022deformation}, our new model presents an explanation of shear banding physics, which connects the atomic-scale level of non-affine displacements \cite{Cui} to macroscopic shear banding and yielding physics.

It is worth noting that while our numerical simulation probes athermal conditions, there is experimental evidence that shear bands have a thermally activated component \cite{klaumunzer2010temperature,maass2011propagation}. Despite this caveat, topological defects offer the advantage of being identifiable within the static structure of glasses \cite{wu2023topology,baggioli2023topological}, establishing a direct connection between the metallic glass pre-, during, and post-deformation stages. In this context, they are utilized to characterize the static structure of the deformed glass, mimicking the experimental scenario of shear band formation. This provides an opportunity for future investigations into thermally activated processes within a framework similar to the Peierls-Nabarro model for dislocation dynamics in crystals at finite temperatures \cite{peierls1940size,nabarro1947dislocations}, opening up new pathways for comprehending amorphous plasticity through topological defects.

\section{Conclusions}
This study presents an approach for modelling shear banding physics, connecting atomic scale dynamics to mesoscale density oscillations within the shear band, in terms of well-defined mathematical objects. We introduce a new model that describes the local deformation within the shear band and the surrounding matrix by a sequential arrangement of alternating topological charges (vortex and anti-vortex topological defects) \cite{baggioli2021plasticity,wu2023topology}. Consequently, we obtain a microscopic displacement field that, when combined with continuum mechanics, provides an accurate analytical solution for the density profile along the shear band that is in excellent agreement with experimental data. This solution not only addresses the observed asymmetry in density variations but also has the potential to elucidate a broader spectrum of
observations that show no apparent densification, that is exclusive dilation.

\section*{Acknowledgments} 
A.Z. gratefully acknowledges funding from the European Union through Horizon Europe ERC Grant number: 101043968 ``Multimech'' and from the Nieders{\"a}chsische Akademie der Wissenschaften zu G{\"o}ttingen in the frame of the Gauss Professorship program. A.Z. and A.B. gratefully acknowledge funding from US Army Research Office through contract nr. W911NF-22-2-0256. Useful discussions with T.W. Sirk, Amelia Y. Liu, M. Baggioli, and T. Petersen are gratefully acknowledged.

 

\end{document}